\documentclass[journal]{IEEEtran}
\usepackage{cite}
\usepackage{algorithmic}
\usepackage{array}
\usepackage[caption=false,font=normalsize,labelfont=sf,textfont=sf]{subfig}
\usepackage{textcomp}
\usepackage{graphicx}
\usepackage{url}
\usepackage{verbatim}
\usepackage{balance}
\usepackage{float}
\usepackage{amsthm}
\usepackage{microtype}
\usepackage{booktabs}
\usepackage{algorithm}
\usepackage[hidelinks]{hyperref}
\usepackage{siunitx}
\usepackage{amsmath,amssymb,amsfonts}
\usepackage{cleveref}
\usepackage{mathrsfs}

\usepackage{bm}

\newcommand{\Input}{\item[\textbf{Input:}]}        
\newcommand{\Output}{\item[\textbf{Output:}]}      

\begin{document}

\title{UACS: Uncertainty-Aware Clustering and Sparse Parameter Estimation for Multi-Condition Wheel Center Force}



\author{Le Wang and Mian Li

\thanks{This work was supported by the National Natural Science Foundation of China under Grant 52275263.}
\thanks{Le Wang is with Division of Decision and Control Systems, School of Electrical Engineering and Computer Science, KTH Royal Institute of Technology, 10044 Stockholm, Sweden. Email: le6@kth.se. Mian Li is with Global Institute of Future Technology, Shanghai Jiao Tong University, 200240 Shanghai, China. Email: mianli@sjtu.edu.cn.}}

\maketitle
\begin{abstract}

Accurate wheel center force (WCF) estimation is important for vehicle durability analysis and condition monitoring, whereas direct measurement requires dedicated transducers and increases instrumentation cost and complexity. As a result, soft sensors have attracted considerable attention as a practical alternative. However, developing reliable soft sensors remains challenging due to varying working conditions, unknown condition relationships, and high-dimensional model representations. This paper proposes an uncertainty-aware clustering and sparse parameter estimation framework (UACS) for WCF signal estimation under diverse and unlabeled working conditions. A MISO-FIR model is first constructed using easily accessible sensor signals to estimate the unmeasurable force signal. The uncertainty of estimated parameters is quantified through confidence ellipsoids, and pairwise hypothesis testing is developed to identify statistically similar working conditions. Based on these uncertainty-aware comparisons, working conditions are grouped in an unsupervised manner, and sparse parameter estimation is performed within each cluster. The UACS framework simultaneously reduces the number of required models and active parameters while maintaining estimation accuracy. Experiments using real-world vehicle data demonstrate that the UACS framework achieves estimation accuracy comparable to condition-specific models while substantially reducing the number of required models and outperforming conventional distance-based clustering approaches.

\end{abstract}


\IEEEpeerreviewmaketitle

\section{Introduction}

\begin{figure}[htbp]
    \centering
    \includegraphics[page=1, width=0.9\linewidth]{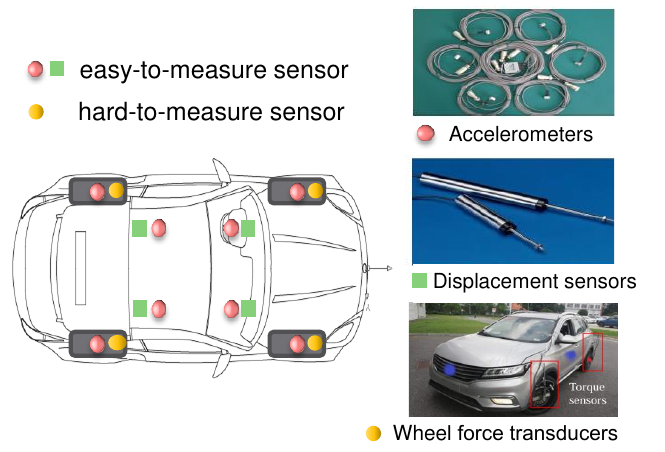}
    \caption{Hard-to-measure and easy-to-measure sensors installed on the vehicle.}
    \label{fig:input}
\end{figure}

\IEEEPARstart{A}{ccurate} and reliable wheel center force (WCF) estimation is important for durability analysis and condition monitoring of vehicle components~\cite{johannesson2013guide,wang2024merging}. Reliable WCF information enables load monitoring, durability assessment, and fault diagnosis under varying working conditions~\cite{na2017vehicle,wang2023hybrid}. However, direct measurement of WCFs in practical applications is challenging because dedicated wheel force transducers increase hardware cost and system complexity and require careful installation, calibration, and maintenance~\cite{taghirad2002intelligent,webster2018measurement}. These limitations motivate the development of data-driven soft sensors that estimate difficult-to-measure physical quantities from inexpensive and readily available sensor measurements~\cite{capriglione2020soft}. An illustrative example is shown in Fig.~\ref{fig:input}, where the test vehicle is equipped with readily available sensors, including accelerometers and suspension displacement sensors, while dedicated wheel force transducers are installed to provide reference measurements. This configuration enables the collection of synchronized onboard sensor signals and reference WCF measurements, which are essential for data-driven model development and validation.

A soft sensor estimates a difficult-to-measure quantity from other readily available measurements using a mathematical or data-driven model. In this work, WCF is estimated from onboard sensor signals using a data-driven multiple-input single-output (MISO) finite impulse response (FIR) model. Multiple measured sensor signals are used as model inputs, while WCF is treated as the output. The resulting linear model leads to a standard regression formulation within the system identification framework~\cite{ljung1998system}.

The MISO-FIR model can be estimated using least squares~(LS), a standard approach for linear parameter estimation due to its simplicity and numerical stability~\cite{ljung1998system,verhaegen2007filtering,kay1993statistical}. In practical soft-sensing applications, variations in working conditions can lead to changes in the underlying input-output relationships and affect model performance~\cite{fang2021soft}. In real-world vehicle operation, such variations arise from different road surfaces and driving conditions. For example, washboard roads, Belgian block roads, resonance roads, speed bumps, and pothole surfaces can produce distinct dynamic responses~\cite{park2014analysis}. Consequently, a single model may not adequately represent all working conditions. Although separate condition-specific models can be constructed, the number of required models increases with the number of working conditions, which complicates model storage, maintenance, and deployment.

This motivates model sharing among working conditions with similar input-output relationships. Sharing models can reduce the number of condition-specific models, but requires a reliable criterion for determining which conditions can share a common model. A natural approach is to compare the parameters estimated for individual conditions. However, these estimates are obtained from noisy measurements and are therefore subject to estimation uncertainty. Consequently, similarity assessment based solely on parameter point estimates may lead to unreliable model-sharing decisions~\cite{casella2024statistical}. Model similarity should therefore be assessed by accounting for the uncertainty associated with the parameter estimates.

Beyond assessing model similarity, the grouping structure among working conditions is generally unknown in advance. Conditions that can share common models must therefore be identified directly from the data without predefined group labels or a prescribed number of groups. This constitutes an unsupervised clustering problem. Representative clustering methods include distance-based approaches such as K-means~\cite{mcqueen1967some,bishop2006pattern}, probabilistic methods such as Gaussian mixture models~\cite{reynolds2009gaussian,viroli2019deep,dempster1977maximum}, graph-based methods such as spectral clustering~\cite{von2007tutorial,ng2001spectral}, and density-based methods such as DBSCAN~\cite{schubert2017dbscan,ester1996density}. In the present problem, the clustering procedure should exploit model-similarity information that accounts for parameter estimation uncertainty, so that the resulting groups are suitable for model sharing.

Even after similar working conditions are grouped and represented by shared models, the resulting MISO-FIR models may still be unnecessarily complex. Because the model contains multiple inputs and a relatively high order, its parameter vector is high-dimensional, while only a subset of parameters may contribute substantially to WCF estimation within a given condition group. Model interpretability and reliable performance under varying conditions are important considerations in practical soft-sensing
applications~\cite{cao2025comprehensive}. This motivates sparse parameter
estimation to further reduce model complexity and improve interpretability. Lasso~\cite{tibshirani1996regression} and its extensions, including group Lasso~\cite{gao2021autonomous,friedman2010note,zhang2019feature,Yuan:06,yuan2011efficient}, sparse group Lasso~\cite{simon2013sparse,rao2015classification,vincent2014sparse}, and fused Lasso~\cite{tibshirani2005sparsity,rinaldo2009properties,cui2021fused}, provide widely used tools for sparsity-promoting estimation. However, these methods alone do not address the multi-condition modeling problem considered here, where both the
number of shared models across working conditions and the number of active parameters within each model need to be reduced.

Taken together, these observations suggest that reliable multi-condition WCF estimation requires addressing uncertainty in model comparison, unknown condition relationships, and parameter redundancy in high-dimensional models. To address these issues, we develop an uncertainty-aware clustering and sparse parameter estimation~(UACS) framework that reduces both the number of required models and the number of active parameters. The key strengths of this framework are as follows.

\begin{itemize}
\item Parameter estimation uncertainty is incorporated into the comparison of condition-specific models through covariance information and statistical hypothesis testing. This enables differences between condition-specific models to be assessed relative to their estimation uncertainty rather than from parameter point estimates alone.

\item The uncertainty-aware comparison is used to identify groups of working conditions without prior knowledge of the grouping structure or the number of groups, enabling statistically similar conditions to share a common model.

\item For each identified group, a common sparse model is estimated from the combined data. This further reduces the number of active parameters while maintaining predictive performance, thereby achieving both model sharing and parameter sparsity.

\item Real-world vehicle experiments and ablation studies demonstrate that UACS reduces the number of required models from 13 to 5 while increasing the average validation MSE by only 2.57\% compared with condition-specific modeling. Sparse estimation further reduces the number of active parameters by 24.7\%--50.0\% across the identified groups. In addition, UACS achieves an approximately 43.4\% reduction in average MSE compared with K-means-based grouping, highlighting the benefit of incorporating parameter uncertainty into condition grouping.
\end{itemize}

\noindent {\bf Notations}: $[K]$ represents the set $\{1,\dots,K\}$ for any positive integer $K$. For a set $S$, let $|S|$ be the cardinality of $S$. Let $x \sim \mathcal{N}(\mu, \Sigma)$ denote a multivariate Gaussian distribution with mean vector $\mu$ and covariance matrix $\Sigma$. Let \(\chi_d^2\) denote a chi-squared random variable with \(d\) degrees of freedom, and let \(\chi^2_{d,q}\) denote its \(q\)-quantile. \( \mathbf{1}(\cdot) \) is the indicator function. For a matrix $A=[A_1,\dots,A_p]\in\mathbb{R}^{n\times p}$ with columns $A_1,\dots,A_p\in\mathbb{R}^p$ and an index set $S=\{i_1,\dots,i_k\}\subset [p]$, let $A_S=[A_{i_1},\dots,A_{i_k}]$. In other words, $A_S$ extracts the columns of $A$ in $S$. Similarly, for a vector $b=(b_1,\dots,b_p)^\top \in\mathbb{R}^n$, let $b_S=(b_{i_1},\dots,b_{i_k})^\top$. In other words, $b_S$ extracts the components of $b$ in $S$. Let $A_{-S}$ and $b_{-S}$ to extract the columns or components not in $S$. Let $\text{sign}(x)$ be the sign function: $\text{sign}(x)=1$ if $x>0$; $\text{sign}(x)=0$ if $x=0$; $\text{sign}(x)=-1$ if $x<0$. Without loss of generality, we also use $\text{sign}(b)$ to denote the sign vector of $b$. Let $\mathcal{A}_b=\{j:\text{sign}(b)_j\neq0\}$ be the active set of $b$.

The remainder of this paper is organized as follows. Section~II presents the problem formulation. Section~III describes the proposed methodology. Section~IV presents real-world experiments. Section~V concludes the paper.

\begin{figure*}[t]
    \centering
    \includegraphics[width=0.98\linewidth]{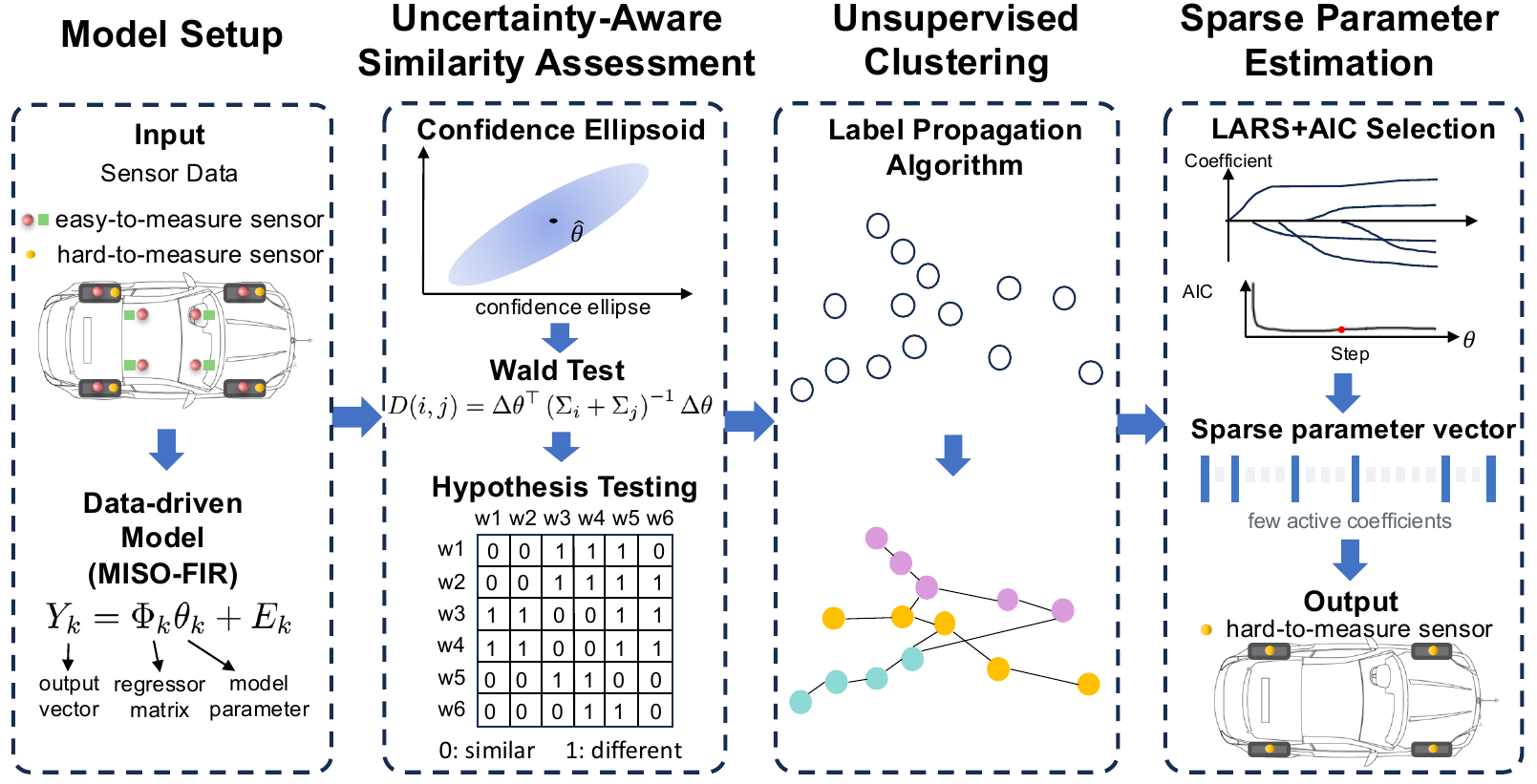}
    \caption{Overview of the proposed UACS framework, consisting of four components: model setup, uncertainty-aware similarity assessment, unsupervised clustering, and sparse parameter estimation.}
    \label{fig:method}
\end{figure*}

\section{Problem Statement}

The problem considered in this paper is to estimate hard-to-measure WCF from easily accessible sensor measurements under multiple working conditions. In the vehicle system studied here, the hard-to-measure signal is the WCF, whereas the easily accessible signals are standard sensor measurements such as accelerations and suspension displacements. 

Vehicles operate under different road surfaces and driving environments, which are treated as different working conditions in this paper. For each working condition, the relationship between the measured sensor
inputs and the WCF output is described by a separate model. Specifically, for working condition $k \in [K]$, the WCF signal is modeled as
\begin{equation}
Y_k = M_{\theta_k}(\Phi_k),
\label{eq:input}
\end{equation}
where $Y_k \in \mathbb{R}^{N_k}$ denotes the WCF output, $\Phi_k \in \mathbb{R}^{N_k \times p}$ is the matrix constructed from the sensor measurement inputs, and $M_{\theta_k}$ is a model parameterized by $\theta_k \in \mathbb{R}^{p}$.

Different working conditions may induce different input-output relationships between the sensor measurements and the WCF signal. As a result, the parameter vectors $\{\theta_k\}_{k=1}^K$ are generally different across conditions. However, some conditions may exhibit similar system behavior and can therefore be grouped and represented by a common parameter vector. For example, the Belgian block and resonance road conditions are similar. In practice, the grouping structure among working conditions is unknown, and it is not clear which conditions can be described by the same parameter vector. A straightforward approach is therefore to estimate a separate model parameter vector $\theta_k$ for each working condition. When the number of working conditions $K$ is large, this leads to a large collection of condition-specific model parameter vectors $\{\theta_k\}_{k=1}^K$, which is inefficient in terms of storage, maintenance, and deployment, especially for embedded implementations. This motivates the need to construct a reduced set of shared models that can represent multiple working conditions while maintaining estimation accuracy.

This problem is challenging for three reasons. First, for each working condition, the parameter vector $\theta_k$ is estimated from a finite number of noisy measurements $(\Phi_k, Y_k)$, which leads to uncertainty in the estimated parameters. Since condition grouping relies on comparing model parameters across working conditions, using point estimates alone may lead to unreliable grouping decisions. Second, although some working conditions may share a common model, the underlying grouping structure is unknown a priori. Therefore, the underlying grouping structure must be inferred directly from the data in an unsupervised way. Third, the parameter vector $\theta_k \in \mathbb{R}^p$ is high-dimensional, while only a subset of its components may be relevant for WCF estimation under a given working condition. Directly using full parameter vectors may therefore introduce redundant parameters and lead to less compact models.

From a mathematical perspective, given condition-specific datasets $\{(\Phi_k, Y_k)\}_{k=1}^K$, the objective is to identify a partition $\{\mathcal{C}_i\}_{i=1}^{K_1}$ with $K_1 \ll K$ and, for each group $\mathcal{C}_i$, determine a shared parameter vector that provides accurate WCF estimation for all $k \in \mathcal{C}_i$. In addition, the shared parameter vector in each group is expected to be sparse, so that only the relevant parameters remain active.

Therefore, the problem considered in this paper is to jointly determine which working conditions can share a common model and how to obtain the corresponding sparse shared models from noisy multi-condition data without prior knowledge of the grouping structure. The overall goal is to achieve accurate WCF estimation across diverse working conditions while reducing both the number of models and the number of active model parameters.

\section{Methodology}
\label{method}
This section presents the UACS framework for WCF estimation under multiple working conditions. As shown in Fig.~\ref{fig:method}, the proposed UACS framework consists of four components: model setup, uncertainty-aware similarity assessment, unsupervised clustering, and sparse parameter estimation. The model setup constructs a data-driven MISO-FIR model for each working condition. The uncertainty-aware similarity assessment component represents parameter uncertainty using confidence ellipsoids and performs hypothesis testing to assess similarity between models. The resulting similarity structure is then used by the label propagation algorithm (LPA) algorithm to cluster working conditions without requiring prior labels \cite{garza2019community}. Within each cluster, sparse parameter estimation is performed using Lasso with least angle regression (LARS) \cite{efron2004least} and Akaike's information criterion (AIC) \cite{cavanaugh2019akaike} to identify relevant parameters and thus reduce model complexity. Each component is described in detail in the following.

\subsection{Model Setup}

We specialize the input-output model \eqref{eq:input} in Section II to WCF estimation using a MISO-FIR model. For the $k$-th working condition, the output is modeled as a linear combination of the current and past input signals~\cite{ljung1998system}:
\begin{align}
\nonumber
y_k(t)
=
a_{0,k}^{\top}x_k(t)
+\cdots+
a_{n,k}^{\top}x_k(t-n)
+e_k(t),\\
\quad t=n+1,\ldots,N_k+n,
\label{equ:linear}
\end{align}
where $n$ is the model order, $N_k$ is the number of output samples under the $k$-th working condition, $y_k(t)\in\mathbb{R}$ is the output, $x_k(t)\in\mathbb{R}^m$ is the model input, $a_{i,k}\in\mathbb{R}^m$, $i=0,\ldots,n$, are the unknown parameters to be estimated, and $e_k(t)$ is measurement noise.

Equation~\eqref{equ:linear} can be written compactly as
\begin{align}
Y_k=\Phi_k\theta_k+E_k,
\label{equ:vector1}
\end{align}
where the model parameter vector is defined as
\[
\theta_k
=
[a_{0,k}^{\top},\ldots,a_{n,k}^{\top}]^{\top}
\in
\mathbb{R}^{p},
\qquad
p=m(n+1),
\]
and
\begin{align*}
Y_k
&=
[y_k(n+1),\ldots,y_k(N_k+n)]^{\top},\\
\Phi_k
&=
\begin{bmatrix}
\phi_k(n+1)^{\top} \\
\vdots \\
\phi_k(N_k+n)^{\top}
\end{bmatrix},\\
E_k
&=
[e_k(n+1),\ldots,e_k(N_k+n)]^{\top},
\end{align*}
with the regressor vector
\[
\phi_k(t)
=
[x_k^{\top}(t),\ldots,x_k^{\top}(t-n)]^{\top}.
\]
Same as the standard assumption in the literature, we assume the noise is white, i.e., $E_k\sim\mathcal{N}(\bm{0},\beta_k I)$, where $\beta_k$ is the
variance.

Applying the LS estimate to each working condition yields a collection of condition-specific parameter estimates $\{\hat{\theta}_k\}$. As mentioned before, when the number of working conditions increases, maintaining a separate model for each condition becomes computationally costly and leads to considerable model redundancy. Moreover, each $\hat{\theta}_k$ is uncertainty due to the measurement noise. The proposed UACS framework addresses this issue by performing uncertainty-aware similarity assessment, clustering statistically similar working conditions, and estimating sparse models for the resulting groups.

\subsection{Uncertainty-Aware Similarity Assessment}

To account for uncertainty in parameter estimation, we first compute the LS estimate for each working condition and characterize its uncertainty through the covariance matrix and the associated confidence ellipsoid. Based on these uncertainty descriptions, we then assess whether two working conditions can be regarded as similar. This step provides the pairwise similarity information used in the subsequent clustering procedure.

We first characterize parameter uncertainty using confidence ellipsoids. For clarity, we denote the estimated parameter vector for the $k$-th working condition $W_k$ by $\hat{\theta}_k$.

\subsubsection{Confidence Ellipsoids}

For each working condition, the model parameters are estimated independently using the LS method. We assume that the regression matrix $\Phi_k \in \mathbb{R}^{N_k \times p}$ has full column rank, i.e.,
\begin{align*}
    \operatorname{rank}(\Phi_k)=p,
\end{align*}
such that $\Phi_k^\top\Phi_k$ is invertible, since the number of samples $N_k$ is typically much larger than the number of parameters $p$. The LS estimate is then given by~\cite{ljung1998system}
\begin{align}
    \hat{\theta}_k
    =
    \left(\Phi_k^\top\Phi_k\right)^{-1}
    \Phi_k^\top Y_k.
    \label{eq:ls}
\end{align}

Due to noisy measurements, the LS estimate is subject to uncertainty. Specifically, substituting \eqref{equ:vector1} into \eqref{eq:ls} gives
\begin{align}
    \hat{\theta}_k
    =
    \theta_k
    +
    \left(\Phi_k^\top\Phi_k\right)^{-1}
    \Phi_k^\top E_k,
\nonumber
\end{align}
which indicates that the estimation error is induced by the measurement noise. Therefore, $\hat{\theta}_k$ is a random variable whose uncertainty can be characterized by its covariance matrix. Under the Gaussian noise assumption in \eqref{equ:vector1}, the covariance matrix of the LS estimate is given by
\begin{align}
    \Sigma_{\hat{\theta}_k}
    =
    \beta_k
    \left(\Phi_k^\top\Phi_k\right)^{-1},
\nonumber
\end{align}
where $\beta_k$ denotes the noise variance. Since $\beta_k$ is unknown, it is estimated from the LS residuals as
\begin{align*}
    \hat{\beta}_k
    =
    \frac{
        \left\|Y_k-\Phi_k\hat{\theta}_k\right\|_2^2
    }{
        N_k-p
    }.
\end{align*}
Therefore, the covariance matrix is estimated by
\begin{align}
    \hat{\Sigma}_{\hat{\theta}_k}
    =
    \hat{\beta}_k
    \left(\Phi_k^\top\Phi_k\right)^{-1}.
        \label{eq:cov}
\end{align}

The covariance matrix characterizes the uncertainty of the parameter estimates by describing the variance of each parameter and the covariance between different parameters. Based on the estimated covariance matrix, a confidence ellipsoid can be constructed. Under the Gaussian noise assumption, the normalized estimation error satisfies
\begin{align*}
    \left(\hat{\theta}_k-\theta_k\right)^\top
    \Sigma_{\hat{\theta}_k}^{-1}
    \left(\hat{\theta}_k-\theta_k\right)
    \sim
    \chi_p^2.
\end{align*}
In practice, $\Sigma_{\hat{\theta}_k}$ is replaced by its estimate $\hat{\Sigma}_{\hat{\theta}_k}$, and the resulting chi-square region is used as an approximation.

Accordingly, an approximate $(1-\alpha)$ confidence ellipsoid is defined as
\begin{align}
    \widehat{\mathcal{C}}_k
    =
    \left\{
    \theta \in \mathbb{R}^p:
    \left(\hat{\theta}_k-\theta\right)^\top
    \hat{\Sigma}_{\hat{\theta}_k}^{-1}
    \left(\hat{\theta}_k-\theta\right)
    \leq
    \chi^2_{p,1-\alpha}
    \right\},
    \label{eq:confidence_ellipsoid}
\end{align}
where $\chi^2_{p,1-\alpha}$ denotes the $(1-\alpha)$-quantile of the chi-square distribution with $p$ degrees of freedom. As mentioned before, when the number of observations in each working condition is substantially larger than the number of estimated parameters, this approximation can used to characterize parameter uncertainty \cite{ljung1998system}.

The confidence ellipsoid provides an uncertainty region for each individual parameter estimate. Therefore, assessing the similarity between two working conditions requires evaluating whether the difference between their parameter estimates is significant relative to their associated uncertainty regions. Specifically, two estimates with a large Euclidean distance do not necessarily indicate statistically different working conditions if the corresponding uncertainties are large. Conversely, even a small distance between two estimates may indicate a significant difference when both estimates have small uncertainties. Therefore, rather than directly comparing the parameter estimates, their difference is evaluated with respect to the joint uncertainty of the two estimates in the subsequent hypothesis testing procedure.

\subsubsection{Hypothesis Testing}

Based on the uncertainty characterization obtained above, we perform pairwise hypothesis testing to determine whether the difference between two parameter estimates is significant relative to their joint uncertainty.

For two working conditions $W_i$ and $W_j$ with parameter vectors $\theta_i$ and $\theta_j$, respectively, the hypotheses are formulated as
\begin{align*}
H_0 &: \theta_i=\theta_j,\\
H_1 &: \theta_i\neq\theta_j.
\end{align*}

We compute the LS estimates of $\theta_i$ and $\theta_j$ using \eqref{eq:ls}. According to the uncertainty characterization in the previous subsection, the estimated covariance matrices of the two estimates are denoted by $\hat{\Sigma}_{\hat{\theta}_i}$ and $\hat{\Sigma}_{\hat{\theta}_j}$, respectively. Under $H_0$, the difference between the two estimates is centered at zero, and its variability is determined by the uncertainties of both estimates. Since the parameter estimates obtained from different working conditions are assumed to be independent, the estimated covariance matrix of their difference is given by
\begin{align*}
\hat{\Sigma}_{\hat{\theta}_i-\hat{\theta}_j}
=
\hat{\Sigma}_{\hat{\theta}_i}
+
\hat{\Sigma}_{\hat{\theta}_j}.
\end{align*}

The Wald statistic is then defined as \cite{wald1943tests}
\begin{align*}
D(i,j)
=
\left(\hat{\theta}_i-\hat{\theta}_j\right)^\top
\left(
\hat{\Sigma}_{\hat{\theta}_i}
+
\hat{\Sigma}_{\hat{\theta}_j}
\right)^{-1}
\left(\hat{\theta}_i-\hat{\theta}_j\right).
\end{align*}

This statistic can be interpreted as the squared Mahalanobis distance between the two parameter estimates, normalized by their joint uncertainty. Under $H_0$, and based on the Gaussian assumption introduced above, $D(i,j)$ approximately follows a chi-square distribution with $p$ degrees of freedom.

The null hypothesis $H_0$ is rejected at significance level $\alpha$ if
\begin{align*}
D(i,j)>\chi^2_{p,1-\alpha},
\end{align*}
indicating that the two working conditions have statistically different parameter vectors. Otherwise,
\begin{align*}
D(i,j)\leq\chi^2_{p,1-\alpha},
\end{align*}
and $H_0$ is not rejected. In this case, no statistically significant difference is detected between the two parameter vectors, and the two working conditions are treated as similar in the subsequent clustering procedure.

Based on the pairwise hypothesis tests, we define a symmetric binary matrix
$R\in\{0,1\}^{K\times K}$, with $R(i,i)=0$. For $i\neq j$,
\begin{align}
R(i,j)
=
\begin{cases}
0, & \text{if } D(i,j)\leq\chi^2_{p,1-\alpha},\\
1, & \text{if } D(i,j)>\chi^2_{p,1-\alpha}.
\end{cases}
\label{eq:rmatrix}
\end{align}

The resulting matrix $R$ records which pairs of working conditions are treated as statistically similar and provides the graph structure used in the subsequent clustering procedure.

\subsection{Unsupervised Clustering}

Based on the pairwise similarity information obtained from hypothesis testing, we cluster the working conditions using LPA \cite{garza2019community}, a graph-based unsupervised clustering method. We first construct a graph from the similarity matrix $R$ and then apply LPA to partition the working conditions into clusters.

\paragraph{Graph Construction}

The binary test-outcome matrix $R$ is interpreted as the adjacency information of an undirected graph $\mathcal{G}=(\mathcal{V},\mathcal{E})$, where each working condition $W_i$ corresponds to a node $v_i\in\mathcal{V}$. According to the definition of $R$, an edge $e(i,j)\in\mathcal{E}$ is added between two distinct nodes $v_i$ and $v_j$ if $R(i,j)=0$, indicating that the corresponding working conditions are treated as statistically similar. Otherwise, no edge is added between them. The neighborhood of node $v_i$ is denoted by
\begin{align*}
\mathcal{N}(v_i)
=
\left\{
v_j\in\mathcal{V}:
j\neq i,\ e(i,j)\in\mathcal{E}
\right\}.
\end{align*}

\paragraph{Initialization}

Initially, each node $v_i$ is assigned a unique label,
\begin{align*}
L(v_i)=i,
\qquad i\in[K],
\end{align*}
so that each working condition is initially treated as an individual cluster.

\paragraph{Iterative Label Propagation}

At each iteration, the nodes are arranged in a random order, denoted by $X$. Their labels are then updated sequentially according to this order:

\begin{itemize}
    \item For each node $v_i$ in $X$, compute the frequency of label $l$ among its neighboring nodes:
    \begin{align*}
    \operatorname{count}_i(l)
    =
    \sum_{v_j\in\mathcal{N}(v_i)}
    \mathbf{1}\bigl(L(v_j)=l\bigr),
    \qquad l\in[K].
    \end{align*}

    \item Define the set of labels with the highest frequency as
    \begin{align*}
    \mathcal{L}_i^{\max}
    =
    \arg\max_{l\in[K]}
    \operatorname{count}_i(l).
    \end{align*}

    \item If $\mathcal{N}(v_i)\neq\emptyset$, update $L(v_i)$ by selecting one label uniformly at random from $\mathcal{L}_i^{\max}$. If $v_i$ has no neighboring nodes, its current label remains unchanged.
\end{itemize}

Since the labels are updated sequentially, the update of a node can use the labels already assigned to previously visited nodes in the current iteration.

\paragraph{Stoping and Clustering}

The label propagation process is repeated until no node label changes during a complete iteration or a predefined maximum number of iterations is reached. After termination, nodes with the same label are assigned to the same cluster.

Let $K_1$ denote the number of distinct labels after termination, and denote these labels in increasing order by $l_1,\ldots,l_{K_1}$. The resulting clusters $\mathcal{C}_1,\ldots,\mathcal{C}_{K_1}$ are defined as
\begin{align}
\mathcal{C}_i
=
\left\{
j\in[K]:
L(v_j)=l_i
\right\},
\qquad i\in[K_1].
\label{eq:cluster}
\end{align}

This approach combines pairwise hypothesis testing with graph-based clustering. The similarity matrix $R$ determines the edges between statistically similar working conditions, while LPA identifies groups according to the resulting graph structure. Because the number of clusters is determined by the propagated labels, it does not need to be specified in advance. The local label-update mechanism also allows the method to be applied efficiently when the number of working conditions is large.

\subsection{Sparse Parameter Estimation}

After clustering, we estimate a common parameter vector for each group of working conditions. To identify the relevant parameters and reduce model redundancy, we perform sparse parameter estimation on the pooled data within each cluster. In the following, we formulate the sparse estimation problem using Lasso, compute the solution path using LARS, and select the final estimate by AIC.

\subsubsection{Lasso}

Within each cluster $\mathcal{C}_i$, data from similar working conditions are pooled to estimate a common parameter vector $\theta$. To identify the relevant parameters and reduce model complexity, we estimate $\theta$ using Lasso:
\begin{align*}
 \min_{\theta} \sum_{k\in\mathcal{C}_i}
 \frac{1}{2}\|Y_k-\Phi_k\theta\|_2^2
 +\lambda\|\theta\|_1,
\end{align*}
where $\lambda>0$ is a regularization parameter that controls the sparsity level.

The corresponding solution path is computed using LARS.

\subsubsection{Least Angle Regression}

The LARS algorithm efficiently computes the entire regularization path of the Lasso problem without repeatedly solving the optimization problem for different values of $\lambda$~\cite{tibshirani2013Lasso,efron2004least}. It traces the solution path by updating the coefficient estimates along a least-angle direction, sequentially adding variables that are most correlated with the residual. Variables may also be removed from the active set when their coefficients shrink to zero. This makes LARS particularly suitable for high-dimensional problems due to its computational efficiency.

For a fixed value of $\lambda$, the Karush--Kuhn--Tucker conditions for the pooled Lasso problem are
\begin{align*}
    \sum_{k\in\mathcal{C}_i}
    \Phi_k^\top(Y_k-\Phi_k\theta)
    =\lambda\gamma,
\end{align*}
where
\begin{align*}
    \gamma_j\in
    \begin{cases}
        \{\operatorname{sign}(\theta_j)\},
        & \text{if } \theta_j\neq 0,\\
        [-1,1],
        & \text{if } \theta_j=0,
    \end{cases}
    \qquad \forall j\in[p],
\end{align*}
and $[p]=\{1,\ldots,p\}$. The equicorrelation set is defined as
\begin{align}
S=
\left\{
j\in[p]:
\left|
\sum_{k\in\mathcal{C}_i}
\Phi_{k,j}^\top(Y_k-\Phi_k\theta)
\right|
=\lambda
\right\},
\label{eq:equicorrelation-set}
\end{align}
and the corresponding equicorrelation signs are
\begin{align}
s=
\operatorname{sign}
\left(
\sum_{k\in\mathcal{C}_i}
\Phi_{k,S}^\top(Y_k-\Phi_k\theta)
\right)
\in\mathbb{R}^{|S|}.
\label{eq:sign-vector}
\end{align}

The LARS algorithm computes the solution path as follows.

\paragraph{Initialization}
Initially, the regularization coefficient $\lambda$ is sufficiently large that all parameter estimates are zero:
\begin{itemize}
    \item Set the iteration counter $l=0$.

    \item Set $\hat{\theta}_i\leftarrow 0$.

    \item Compute the initial correlations
    \begin{align*}
        C_j=
        \sum_{k\in\mathcal{C}_i}
        \Phi_{k,j}^\top Y_k,
        \qquad j\in[p].
    \end{align*}

    \item Determine the first transition point
    \begin{align*}
        \lambda_0=\max_{j\in[p]}|C_j|.
    \end{align*}

    \item Determine the equicorrelation set $S$ by
    \eqref{eq:equicorrelation-set}, i.e.,
    \begin{align*}
        S=\left\{j\in[p]:|C_j|=\lambda_0\right\},
    \end{align*}
    and the corresponding equicorrelation signs $s$ by
    \eqref{eq:sign-vector}, i.e.,
    \begin{align*}
        s=\operatorname{sign}(C_S)
        \in\mathbb{R}^{|S|}.
    \end{align*}
\end{itemize}

\paragraph{While $\lambda_l>0$}
We compute the solution path and decrease $\lambda$ until the next joining time or crossing time, whichever occurs first:

\begin{itemize}
    \item Compute the coefficient solution path
    $\hat{\theta}_i(\lambda)$ on the current path segment as
    \begin{align}
    \begin{cases}
    \hat{\theta}_{i,S}(\lambda)
    =c-\lambda d
    \in\mathbb{R}^{|S|},
    & \lambda\in[\lambda_{l+1},\lambda_l],\\
    \hat{\theta}_{i,-S}(\lambda)
    =0
    \in\mathbb{R}^{p-|S|},
    & \lambda\in[\lambda_{l+1},\lambda_l],
    \end{cases}
    \label{eq:solution-path}
    \end{align}
    where, assuming that the pooled active regressor matrix has full column rank, $c$ and $d$ are given by
    \begin{align*}
    c={}&
    \left(
    \sum_{k\in\mathcal{C}_i}
    \Phi_{k,S}^\top\Phi_{k,S}
    \right)^{-1}
    \sum_{k\in\mathcal{C}_i}
    \Phi_{k,S}^\top Y_k
    \in\mathbb{R}^{|S|},\\
    d={}&
    \left(
    \sum_{k\in\mathcal{C}_i}
    \Phi_{k,S}^\top\Phi_{k,S}
    \right)^{-1}s
    \in\mathbb{R}^{|S|}.
    \end{align*}

   \item Compute the next joining time and coordinate:
\begin{itemize}
    \item For each $j\notin S$, define
    \begin{align*}
    g_j(\lambda)
    =
    \sum_{k\in\mathcal{C}_i}
    \Phi_{k,j}^{\top}
    \left(
    Y_k-\Phi_{k,S}\hat{\theta}_{i,S}(\lambda)
    \right).
    \end{align*}
    The joining time of the $j$-th variable is
    \begin{align*}
    t_j^{\mathrm{join}}
    =
    \max
    \left\{
    \lambda\in[0,\lambda_l):
    |g_j(\lambda)|=\lambda
    \right\}.
    \end{align*}
    By convention, set $t_j^{\mathrm{join}}=0$ if no positive solution
    exists in $[0,\lambda_l)$. If $S=[p]$, set
    $\lambda_{l+1}^{\mathrm{join}}=0$. If $S\neq[p]$, the next joining
    time is
    \begin{align*}
    \lambda_{l+1}^{\mathrm{join}}
    =
    \max_{j\notin S}t_j^{\mathrm{join}}.
    \end{align*}

        \item If $\lambda_{l+1}^{\mathrm{join}}>0$, determine the joining coordinate
        \begin{align*}
        j_{l+1}^{\mathrm{join}}
        =
        \arg\max_{j\notin S}t_j^{\mathrm{join}},
        \end{align*}
        and its sign
        \begin{align*}
        s_{l+1}^{\mathrm{join}}
        =
        \operatorname{sign}
        \left(
        g_{j_{l+1}^{\mathrm{join}}}
        \bigl(\lambda_{l+1}^{\mathrm{join}}\bigr)
        \right).
        \end{align*}
    \end{itemize}

    \item Compute the next crossing time:
    \begin{itemize}
        \item For each $j\in S$, compute the crossing time of the
        $j$-th variable as
        \begin{align*}
        t_j^{\mathrm{cross}}
        =
        \begin{cases}
        \dfrac{c_j}{d_j},
        &
        d_j\neq0
        \text{ and }
        0\leq\dfrac{c_j}{d_j}<\lambda_l,\\[2mm]
        0,
        & \text{otherwise}.
        \end{cases}
        \end{align*}
        The next crossing time is
        \begin{align*}
        \lambda_{l+1}^{\mathrm{cross}}
        =
        \max_{j\in S}t_j^{\mathrm{cross}}.
        \end{align*}

        \item If $\lambda_{l+1}^{\mathrm{cross}}>0$, determine the crossing coordinate
        \begin{align*}
        j_{l+1}^{\mathrm{cross}}
        =
        \arg\max_{j\in S}t_j^{\mathrm{cross}}.
        \end{align*}
    \end{itemize}

    \item Determine the next transition point
    \begin{align}
    \lambda_{l+1}
    =
    \max
    \left\{
    \lambda_{l+1}^{\mathrm{join}},
    \lambda_{l+1}^{\mathrm{cross}}
    \right\}.
    \label{eq:transition}
    \end{align}

    \item Update the coefficient estimate along the current path segment:
    \begin{align*}
        \hat{\theta}_i
        \leftarrow
        \hat{\theta}_i(\lambda_{l+1}).
    \end{align*}

    \item Update the equicorrelation set and signs:
    \begin{itemize}
        \item If
        $\lambda_{l+1}^{\mathrm{join}}>
        \lambda_{l+1}^{\mathrm{cross}}$,
        add the joining coordinate
        $j_{l+1}^{\mathrm{join}}$ to $S$ and its sign
        $s_{l+1}^{\mathrm{join}}$ to $s$.

        \item If
        $\lambda_{l+1}^{\mathrm{join}}<
        \lambda_{l+1}^{\mathrm{cross}}$,
        remove the crossing coordinate
        $j_{l+1}^{\mathrm{cross}}$ and its corresponding sign
        from $S$ and $s$, respectively.

        \item If
        $\lambda_{l+1}^{\mathrm{join}}
        =
        \lambda_{l+1}^{\mathrm{cross}}>0$,
        perform both updates.
    \end{itemize}

    \item Set $l\leftarrow l+1$.
\end{itemize}

\subsubsection{Akaike's Information Criterion}

After computing the Lasso regularization path for each group
$\mathcal{C}_i$ using LARS, we use AIC to select the optimal
regularization parameter and the corresponding sparse parameter estimate.

\paragraph{Variance estimation}
For model selection, a common residual variance is estimated from the pooled data within each cluster. We first use the LS method to obtain the pooled parameter estimate
\begin{align*}
\tilde{\theta}_i
=
\left(
\sum_{k\in\mathcal{C}_i}
\Phi_k^\top\Phi_k
\right)^{-1}
\sum_{k\in\mathcal{C}_i}
\Phi_k^\top Y_k,
\end{align*}
and the corresponding residual variance estimate
\begin{align}
\tilde{\beta}_i
=
\frac{
\displaystyle
\sum_{k\in\mathcal{C}_i}
\|Y_k-\Phi_k\tilde{\theta}_i\|_2^2
}{
\displaystyle
\sum_{k\in\mathcal{C}_i}N_k-p
}.
\label{eq:beta-cluster}
\end{align}

\paragraph{Optimal regularization coefficient and Lasso estimate}
Let
\begin{align*}
\mathcal{A}_i(\lambda)
=
\operatorname{supp}
\bigl(\hat{\theta}_i(\lambda)\bigr)
=
\left\{
j\in[p]:
\hat{\theta}_{i,j}(\lambda)\neq0
\right\}
\end{align*}
denote the active set of the Lasso estimate. Using
$\tilde{\beta}_i$ as a fixed plug-in estimate of the residual variance,
we select the regularization coefficient using the normalized AIC
criterion~\cite{zou2007degrees}:
\begin{align}
\lambda_i^*
=
\arg\min_{\lambda\in\{\lambda_l\}}
\left\{
\frac{
\displaystyle
\sum_{k\in\mathcal{C}_i}
\|Y_k-\Phi_k\hat{\theta}_i(\lambda)\|_2^2
}{
\displaystyle
\left(
\sum_{k\in\mathcal{C}_i}N_k
\right)\tilde{\beta}_i
}
+
\frac{
2|\mathcal{A}_i(\lambda)|
}{
\displaystyle
\sum_{k\in\mathcal{C}_i}N_k
}
\right\}.
\label{eq:lambdastar}
\end{align}
The resulting sparse parameter estimate for cluster $\mathcal{C}_i$ is
$\hat{\theta}_i(\lambda_i^*)$, as computed by \eqref{eq:solution-path}.

\subsection{Summary}

Algorithm~1 summarizes the main computational steps of the proposed UACS framework. The method starts with condition-specific parameter estimation and uncertainty characterization, followed by similarity assessment, unsupervised clustering, and sparse parameter estimation.

\begin{algorithm}[h]
\caption{The Proposed UACS Framework}
\begin{algorithmic}[1]
\Input Input-output data of the $K$ working conditions, MISO-FIR model order $n$, and significance level $\alpha$.  \\

\textbf{Model Setup:}
\FOR{$k = 1$ to $K$}
    \STATE Form the regression model according to \eqref{equ:vector1}.
\ENDFOR

\textbf{Uncertainty-Aware Similarity Assessment:}
\FOR{$k = 1$ to $K$}
    \STATE Compute the estimated parameter vector $\hat{\theta}_k$ and its covariance matrix $\Sigma_{\hat{\theta}_k}$ using \eqref{eq:ls} and \eqref{eq:cov}, respectively.
\ENDFOR
\STATE Construct the binary test-outcome matrix $R$ based on confidence ellipsoids and pairwise hypothesis testing, as defined in \eqref{eq:rmatrix}.

\textbf{Unsupervised Clustering:}
\STATE Obtain the clusters $\mathcal{C}_1,\dots,\mathcal{C}_{K_1}$ by applying LPA to the similarity matrix $R$, as given in \eqref{eq:cluster}.

\textbf{Sparse Parameter Estimation:}
\FOR{$i = 1$ to $K_1$}
    \STATE Compute the coefficient solution path $\hat{\theta}_i(\lambda)$ and the transition points $\{\lambda_l\}$ using LARS, as given in \eqref{eq:solution-path} and \eqref{eq:transition}, respectively.
    \STATE Compute the estimated variance $\tilde{\beta}_i$ using \eqref{eq:beta-cluster}.
    \STATE Select the optimal regularization coefficient $\lambda_i^*$ by AIC, as given in \eqref{eq:lambdastar}.
\ENDFOR

\Output Clusters $\{\mathcal{C}_i\}_{i=1}^{K_1}$ and the corresponding sparse parameter estimates $\{\hat{\theta}_i(\lambda_i^*)\}_{i=1}^{K_1}$.
\end{algorithmic}
\end{algorithm}

The proposed UACS framework integrates uncertainty-aware similarity assessment, unsupervised clustering, and sparse parameter estimation into a unified pipeline. By incorporating parameter uncertainty through confidence ellipsoids and hypothesis testing, it enables uncertainty-normalized statistical comparisons between working conditions. The clustering step reduces the number of models, while sparse estimation within each cluster further removes redundant parameters and improves interpretability.

Overall, the UACS framework reduces both the number of models and the number of active parameters, leading to an effective and interpretable representation of system dynamics across diverse working conditions.

\section{Experiments and Results}

\subsection{Experimental Setup}

\begin{figure}[htbp!]
    \centering
    \includegraphics[
        width=0.96\linewidth,
        height=3.3cm
    ]{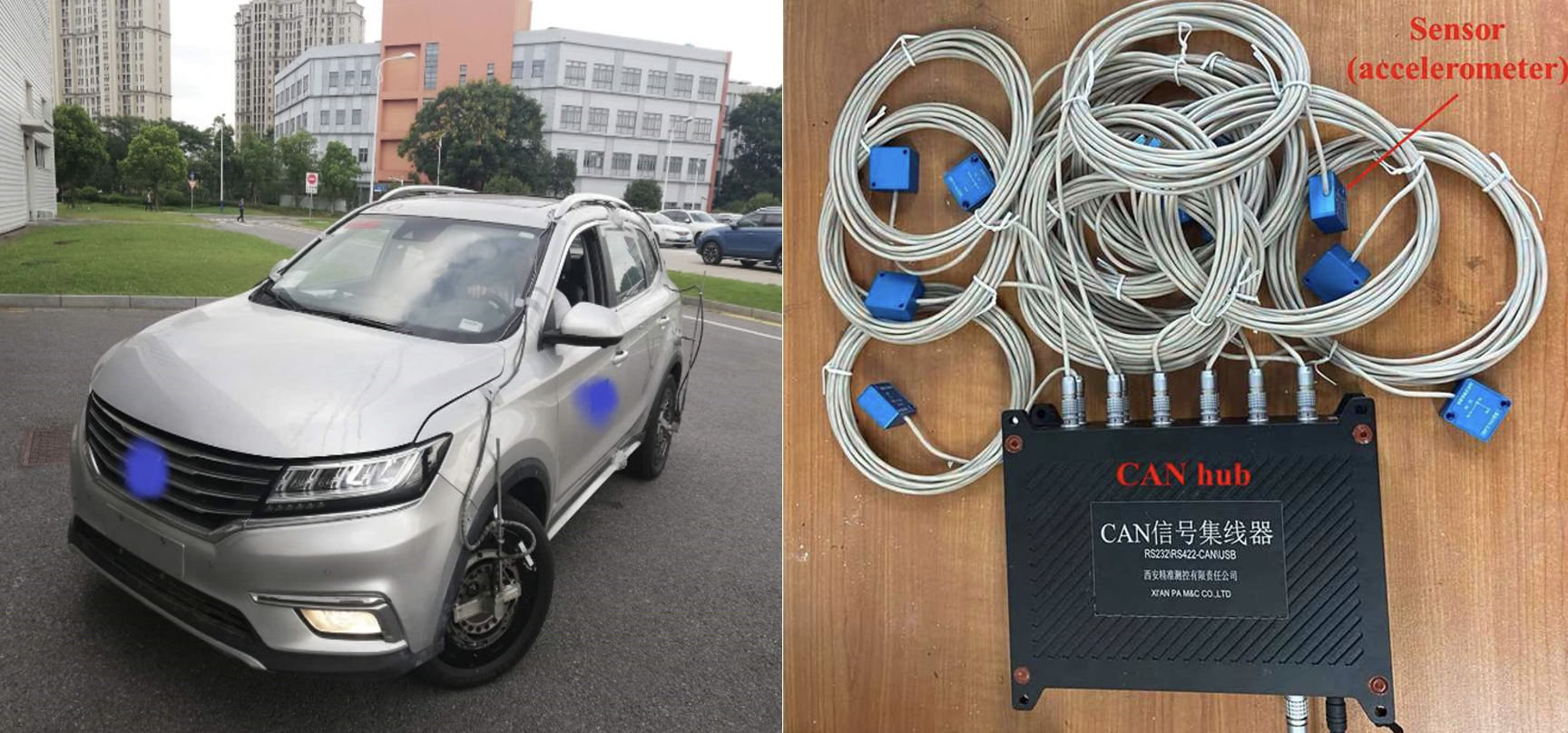}
    \caption{The test vehicle (left) and the CAN hub (right).}
    \label{fig:car}
\end{figure}

\begin{figure}[htbp!]
    \centering
\includegraphics[width=0.96\linewidth]{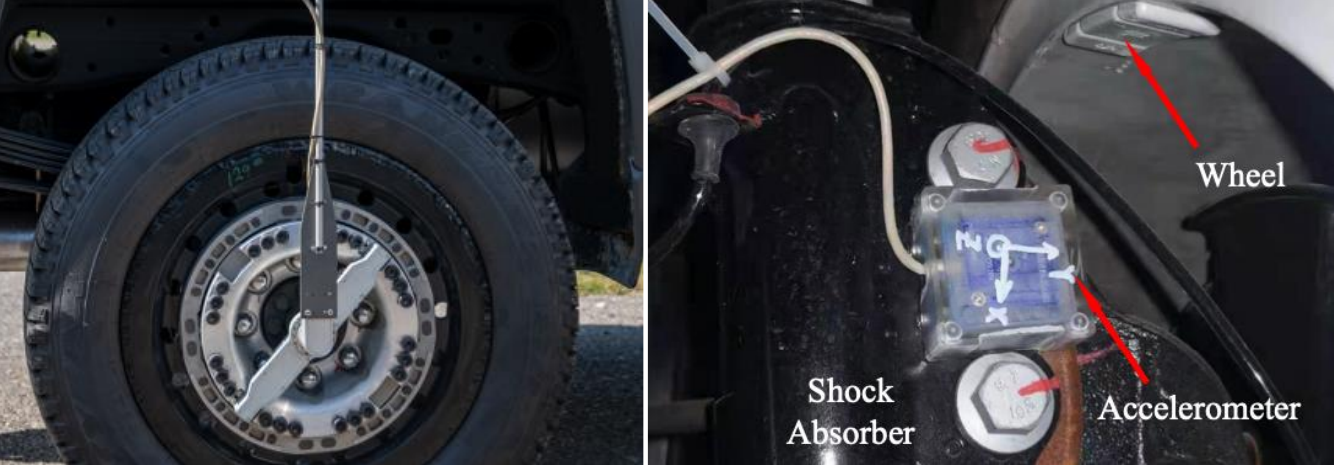}
    \caption{Sensors installed on the test vehicle. Left: The wheel force transducer installed on the test vehicle. Right: The accelerometers (three-axis) installed on the wheel.}
    \label{fig:installation}
\end{figure}

To evaluate the proposed UACS framework, we use experimental data
collected from real-world tests on a prototype part-time four-wheel-drive
vehicle. Fig.~\ref{fig:car} shows the test vehicle and the CAN hub used
for data acquisition. The dataset includes measurements of the WCF and
signals from multiple onboard sensors. The corresponding wheel force
transducer and accelerometer installations are shown in
Fig.~\ref{fig:installation}.

The order of the MISO-FIR model is set to $n=49$. Six sensor signals
are selected as model inputs from a larger sensor pool. Their definitions
are listed in TABLE~\ref{tab:sensor}. The output is the right-front WCF
in the $Z$ direction. Therefore, the dimension of the parameter vector
is
\[
p=(49+1)\times6=300.
\]
Each block of 50 parameters corresponds to one input signal. Before
constructing the regression matrices, the input and output signals are
standardized.

Table~\ref{tab:def} summarizes the 13 working conditions considered
in the experiments. These conditions are selected from a larger set of
available working conditions and include different road-excitation
types, braking maneuvers, and vehicle speeds. Separate training and
validation datasets are used for each working condition.

\begin{table}[t]
  \centering
  \renewcommand{\arraystretch}{1.2}
  \caption{Definitions of the input sensors and output sensor.}
  \begin{tabular}{ccc}
    \toprule
    & Sensor & Category \\
    & Number & \\
    \midrule
    &57 & Throttle pedal\\
    &70 & Engine speed\\
    Input&91 & Right-front suspension displacement\\
    Sensors&92 & Left-front suspension displacement\\
    &93 & Left-rear suspension displacement\\
    &94 & Right-rear suspension displacement\\
    \midrule
    Output&41 & Right-front WCF in the $Z$ direction\\
    \bottomrule
  \end{tabular}
  \label{tab:sensor}
\end{table}

\begin{table}[t]
    \centering
    \caption{Definitions of the 13 working conditions.}
    \renewcommand{\arraystretch}{1.2}
    \setlength{\tabcolsep}{5pt}
    \begin{tabular}{lll}
        \toprule
        No. & Working & Definition\\
        & Condition &\\
        \midrule
        1 & BR30  & Belgian block road at 30 km/h\\
        2 & BR40  & Belgian block road at 40 km/h\\
        3 & RE30  & Resonance road at 30 km/h\\
        4 & RE40  & Resonance road at 40 km/h\\
        5 & WB40  & Washboard with variable distance at 40 km/h\\
        6 & WBA40 & Washboard asynchronous at 40 km/h\\
        7 & SW20  & Short-wave road at 20 km/h\\
        8 & SW25  & Short-wave road at 25 km/h\\
        9 & BK30  & Braking at 30 km/h\\
        10 & BK50 & Braking at 50 km/h\\
        11 & BK70 & Braking at 70 km/h\\
        12 & CJ30 & Cement-joint road at 30 km/h\\
        13 & CJ50 & Cement-joint road at 50 km/h\\
        \bottomrule
    \end{tabular}
    \label{tab:def}
\end{table}

\subsection{Results of Uncertainty-Aware Similarity Assessment}

\begin{table*}[t]
    \centering
    \caption{Binary test-outcome matrix $R$.}
    \renewcommand{\arraystretch}{1.2}
    \setlength{\tabcolsep}{2pt}
    \begin{tabular}{c*{13}{c}}
    \toprule
        & BR30 & BR40 & RE30 & RE40 & WB40 & WBA40
        & SW20 & SW25 & BK30 & BK50 & BK70 & CJ30 & CJ50 \\
    \midrule
    BR30  & 0 & 0 & 0 & 1 & 1 & 1 & 1 & 1 & 1 & 1 & 1 & 1 & 1 \\
    BR40  & 0 & 0 & 0 & 0 & 1 & 1 & 1 & 1 & 1 & 1 & 1 & 1 & 1 \\
    RE30  & 0 & 0 & 0 & 1 & 1 & 1 & 1 & 1 & 1 & 1 & 1 & 1 & 1 \\
    RE40  & 1 & 0 & 1 & 0 & 1 & 1 & 1 & 1 & 1 & 1 & 1 & 1 & 1 \\
    WB40  & 1 & 1 & 1 & 1 & 0 & 0 & 1 & 1 & 1 & 1 & 1 & 1 & 1 \\
    WBA40 & 1 & 1 & 1 & 1 & 0 & 0 & 1 & 1 & 1 & 1 & 1 & 1 & 1 \\
    SW20  & 1 & 1 & 1 & 1 & 1 & 1 & 0 & 0 & 1 & 1 & 1 & 1 & 1 \\
    SW25  & 1 & 1 & 1 & 1 & 1 & 1 & 0 & 0 & 1 & 1 & 1 & 1 & 1 \\
    BK30  & 1 & 1 & 1 & 1 & 1 & 1 & 1 & 1 & 0 & 0 & 0 & 1 & 1 \\
    BK50  & 1 & 1 & 1 & 1 & 1 & 1 & 1 & 1 & 0 & 0 & 0 & 1 & 1 \\
    BK70  & 1 & 1 & 1 & 1 & 1 & 1 & 1 & 1 & 0 & 0 & 0 & 1 & 1 \\
    CJ30  & 1 & 1 & 1 & 1 & 1 & 1 & 1 & 1 & 1 & 1 & 1 & 0 & 0 \\
    CJ50  & 1 & 1 & 1 & 1 & 1 & 1 & 1 & 1 & 1 & 1 & 1 & 0 & 0 \\
    \bottomrule
    \end{tabular}
    \label{tab:hy}
\end{table*}

Table~\ref{tab:hy} presents the binary test-outcome matrix obtained
using $\alpha=0.05$. Here, $R(i,j)=0$ indicates that no statistically
significant parameter difference is detected between two working
conditions, whereas $R(i,j)=1$ indicates that the null hypothesis is
rejected.

Several block-like patterns of zeros are observed. In particular, no
significant pairwise differences are detected within the groups
SW20--SW25, WB40--WBA40, BK30--BK50--BK70, and CJ30--CJ50.
BR30, BR40, and RE30 are also mutually connected, while RE40 is
connected to BR40. These pairwise relations define the graph used in
the subsequent LPA clustering procedure.

\subsection{Sensitivity to the Significance Level}

To examine the influence of the significance level, the pairwise
testing and LPA procedures are repeated using
$\alpha=0.01$, $0.05$, and $0.10$. The same five-cluster partition
is obtained for all three values. Thus, the clustering result remains
unchanged over the considered range of significance levels.

\subsection{Clustering Results and Comparison}

\begin{figure*}[t]
    \centering

    \subfloat[
    {\rmfamily\footnotesize Clustering result obtained by LPA.}
    \label{fig:lpa}]
    {
        \includegraphics[
            width=0.31\textwidth,
            height=4.1cm
        ]{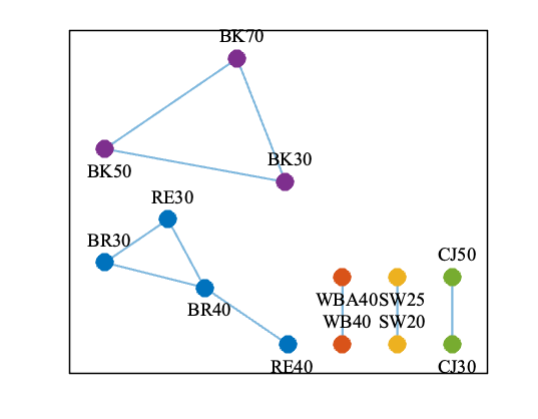}
    }
    \hfill
    \subfloat[
    {\rmfamily\footnotesize Parameter entry steps along the LARS
    solution path for the SW20--SW25 cluster.}
    \label{fig:lars}]
    {
        \includegraphics[
            width=0.31\textwidth,
            height=4.1cm
        ]{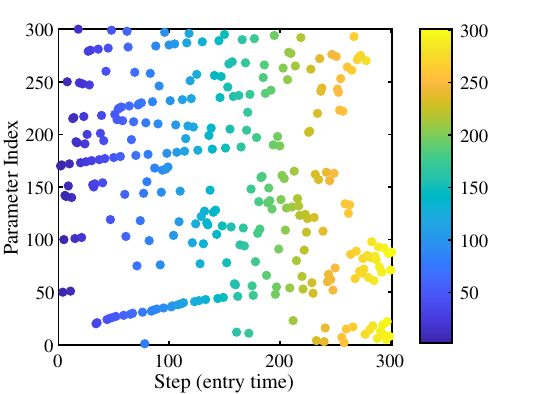}
    }
    \hfill
    \subfloat[
    {\rmfamily\footnotesize Measured and predicted WCF under the
    SW20 working condition.}
    \label{fig:output}]
    {
        \includegraphics[
            width=0.31\textwidth,
            height=4.1cm
        ]{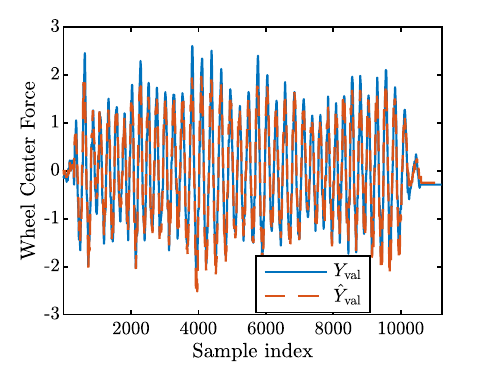}
    }

    \par\vspace{2mm}

    \subfloat[
    {\rmfamily\footnotesize Condition-specific LS parameter estimates
    for SW20 and SW25 and the shared sparse parameter estimate obtained
    by UACS for the SW20--SW25 cluster.}
    \label{fig:theta}]
    {
        \includegraphics[
            width=0.96\textwidth
        ]{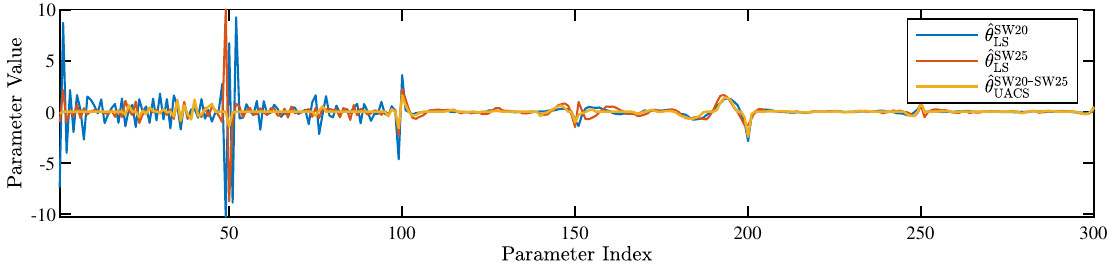}
    }

    \caption{Clustering and sparse parameter estimation results:
    (a) LPA clustering result;
    (b) parameter entry steps along the LARS solution path;
    (c) WCF estimation under SW20; and
    (d) comparison of the condition-specific LS estimates and the
    shared sparse UACS estimate.}
    \label{fig:uacs_results}
\end{figure*}

LPA partitions the 13 working conditions into 5 clusters, as shown
in Fig.~\ref{fig:lpa} and TABLE~\ref{tab:cluster}.

The clustering result is consistent with several operating
characteristics of the working conditions. SW20 and SW25 are grouped
as short-wave road conditions; BK30, BK50, and BK70 are grouped as
braking conditions; CJ30 and CJ50 are grouped as cement-joint road
conditions; and WB40 and WBA40 are grouped as washboard-type
conditions. BR30, BR40, RE30, and RE40 are also assigned to the same
cluster. Although RE40 is not directly connected to BR30 or RE30,
these four conditions form a connected local structure through BR40,
allowing LPA to propagate a common label across the corresponding
graph component.

For comparison, K-means is applied to the condition-specific LS
parameter estimates. The number of K-means clusters is selected as six
using the elbow method \cite{liu2020determine}. Unlike UACS, K-means uses Euclidean distances
between parameter point estimates and does not account for the
uncertainty associated with the estimated parameters.

The K-means partition differs from the UACS result. In particular,
SW20 and SW25 are assigned to different clusters, BK70 is separated
from BK30 and BK50, and CJ30 and CJ50 are also separated. Moreover,
WB40 and WBA40 are grouped with the BR and RE conditions. These
differences illustrate how clustering based only on point-estimate
distances can produce a grouping that differs from the structure
identified using uncertainty-normalized pairwise tests.

\begin{table}[ht]
  \centering
  \renewcommand{\arraystretch}{1.2}
  \caption{Clustering results for the working conditions.}
  \begin{tabular}{ccc}
    \toprule
    Working & Cluster & Cluster \\
    Condition & (UACS) & (K-means) \\
    \midrule
    BR30  & 1 & 1 \\
    BR40  & 1 & 1 \\
    RE30  & 1 & 1 \\
    RE40  & 1 & 1 \\
    WB40  & 2 & 1 \\
    WBA40 & 2 & 1 \\
    SW20  & 3 & 2 \\
    SW25  & 3 & 3 \\
    BK30  & 4 & 2 \\
    BK50  & 4 & 2 \\
    BK70  & 4 & 4 \\
    CJ30  & 5 & 5 \\
    CJ50  & 5 & 6 \\
    \bottomrule
  \end{tabular}
  \label{tab:cluster}
\end{table}

\subsection{Estimation Performance and Ablation Study}

After obtaining the clusters, we evaluate the prediction performance
and model complexity of the resulting models. Mean square error (MSE)
is used as the evaluation metric. For a given working condition, the
MSE is defined as
\[
\mathrm{MSE}
=
\frac{1}{N_{\mathrm{val}}}
\left\|
Y_{\mathrm{val}}
-
\hat{Y}_{\mathrm{val}}
\right\|_2^2,
\]
where $Y_{\mathrm{val}}$ and $\hat{Y}_{\mathrm{val}}$ denote the
measured and predicted validation outputs, respectively, and
$N_{\mathrm{val}}$ is the number of validation samples.

Table~\ref{tab:Lasso} compares the condition-wise validation MSE
obtained using condition-specific LS, the proposed UACS framework, and
K-means-based grouping. For the condition-specific LS baseline, one
parameter vector is estimated separately from the training data of each
working condition. For the proposed UACS framework, one common sparse
parameter vector is estimated from the combined training data within
each UACS cluster. For the K-means-based method, one common LS
parameter vector is estimated from the combined training data within
each K-means cluster.

\begin{table}[t]
    \centering
    \caption{MSE comparison using LS, the proposed method (UACS), and K-means.}
    \renewcommand{\arraystretch}{1.2}
    \setlength{\tabcolsep}{4pt}
    \begin{tabular}{c c c c c}
        \toprule
        Working & MSE & MSE & MSE & $\Delta$ MSE \\
        Condition & (LS) & (UACS) & (K-means) & (UACS-LS) \\
        \midrule
        BR30  & 0.0548 & 0.0625 & 0.0931 & +0.0077 \\
        BR40  & 0.0584 & 0.0598 & 0.0904 & +0.0014 \\
        RE30  & 0.0474 & 0.0608 & 0.0828 & +0.0134 \\
        RE40  & 0.0449 & 0.0546 & 0.0678 & +0.0097 \\
        WB40  & 0.0237 & 0.0250 & 0.2618 & +0.0013 \\
        WBA40 & 0.0676 & 0.0678 & 0.2400 & +0.0002 \\
        SW20  & 0.0383 & 0.0380 & 0.0626 & -0.0003 \\
        SW25  & 0.0682 & 0.0558 & 0.0682 & -0.0124 \\
        BK30  & 0.0435 & 0.0395 & 0.0482 & -0.0040 \\
        BK50  & 0.0424 & 0.0372 & 0.0497 & -0.0052 \\
        BK70  & 0.0181 & 0.0208 & 0.0181 & +0.0027 \\
        CJ30  & 0.1229 & 0.1237 & 0.1229 & +0.0008 \\
        CJ50  & 0.0781 & 0.0808 & 0.0781 & +0.0027 \\
        \bottomrule
    \end{tabular}
    \label{tab:Lasso}
\end{table}

To quantify the difference between UACS and the condition-specific LS
baseline, TABLE~\ref{tab:Lasso} also reports
\[
\Delta\mathrm{MSE}
=
\mathrm{MSE}_{\mathrm{UACS}}
-
\mathrm{MSE}_{\mathrm{LS}}.
\]
The average MSE values of condition-specific LS, UACS, and K-means are
0.0545, 0.0559, and 0.0987, respectively. Thus, UACS reduces the
number of models from 13 to 5 while maintaining an average
prediction error close to that of condition-specific LS.

UACS achieves a lower validation MSE than condition-specific LS for
SW20, SW25, BK30, and BK50. 
The average
MSE of UACS is approximately 2.57\% higher than that of
condition-specific LS, while the number of models is reduced by
\[
\frac{13-5}{13}\times100\%=61.5\%.
\]

Table~\ref{tab:Lasso} also shows that the K-means-based models produce
larger validation errors than UACS for 10 of the 13 working conditions.
The largest differences occur for WB40 and WBA40, which K-means groups
with the BR and RE conditions. The corresponding shared model therefore
has difficulty representing the washboard-type excitation data.

To separate the effect of condition grouping from that of sparse
estimation, an additional ablation experiment is conducted. The UACS
cluster assignments are retained, but sparse estimation is replaced by
LS estimation. Specifically, one common LS parameter vector is
estimated from the combined training data within each UACS cluster.
The results are summarized in TABLE~\ref{tab:overall_comparison}.

\begin{table*}[t]
\centering
\caption{Ablation and overall comparison of different modeling strategies.}
\label{tab:overall_comparison}
\renewcommand{\arraystretch}{1.15}

\resizebox{\textwidth}{!}{
\begin{tabular}{lccc}
\toprule
Modeling strategy
& Estimation within each group
& No. of models
& Average MSE \\
\midrule

Proposed UACS framework
& One shared Lasso model for each UACS cluster
& 5
& 0.0559 \\

Condition-specific modeling
& Separate LS model for each working condition
& 13
& 0.0545 \\

UACS framework without sparse estimation
& One shared LS model for each UACS cluster
& 5
& 0.0560 \\

K-means clustering without sparse estimation
& One shared LS model for each K-means cluster
& 6
& 0.0987 \\
\bottomrule
\end{tabular}
}
\end{table*}
Using one common LS model for each UACS cluster gives an average MSE
of 0.0560, which is close to the 0.0545 obtained using 13 separate
condition-specific LS models. This result shows that the identified
clusters allow multiple working conditions to share a common model
without a substantial increase in average validation error.

When sparse estimation is applied to the same UACS clusters, the
average MSE changes slightly from 0.0560 to 0.0559. This small
difference indicates that the main effect of sparse estimation in the
present experiment is to reduce the number of active coefficients while
preserving the prediction performance of the corresponding shared LS
models.

\begin{table}[t]
\centering
\caption{Numbers of active parameters in the cluster-level models.}
\renewcommand{\arraystretch}{1.15}
\setlength{\tabcolsep}{3pt}
\begin{tabular}{cccc}
\toprule
Cluster & Working conditions & Active & Reduction\\
& & parameters &\\
\midrule
1 & BR30, BR40, RE30, RE40 & 226/300 & 24.7\%\\
2 & WB40, WBA40             & 215/300 & 28.3\%\\
3 & SW20, SW25              & 150/300 & 50.0\%\\
4 & BK30, BK50, BK70        & 200/300 & 33.3\%\\
5 & CJ30, CJ50              & 152/300 & 49.3\%\\
\bottomrule
\end{tabular}
\label{tab:sparse_size}
\end{table}

As shown in TABLE~\ref{tab:sparse_size}, all five cluster-level models
contain fewer active coefficients than the original 300-parameter
model. The reductions range from 24.7\% to 50.0\%. The largest
reductions are obtained for the SW20--SW25 and CJ30--CJ50 clusters,
which retain 150 and 152 active parameters, respectively.

The SW20--SW25 cluster is used to illustrate the sparse estimation
procedure. Fig.~\ref{fig:lars} shows the parameter entry steps along
the LARS solution path. The AIC reaches its minimum at step 151, and
the corresponding parameter vector is selected as the sparse estimate.
As reported in TABLE~\ref{tab:sparse_size}, this estimate retains 150
of the original 300 parameters.

Fig.~\ref{fig:output} compares the
predicted WCF with the measured reference signal under SW20. Fig.~\ref{fig:theta} compares the condition-specific LS estimates for
SW20 and SW25 with the shared sparse estimate for their cluster. The
shared model preserves the main parameter patterns while setting half
of the coefficients to zero. The
predicted response follows the measured signal over the validation
interval, consistent with the MSE reported in TABLE~\ref{tab:Lasso}.

The condition-specific and cluster-level models are estimated using
thousands to tens of thousands of time-series samples and evaluated on
separate validation datasets. Therefore, the reported MSE values
measure prediction performance on data not used for parameter
estimation.

\subsection{Summary}

The experiments show that UACS reduces the number of models from 13
to 5 while increasing the average MSE from 0.0545 to only 0.0559.
The same five-cluster partition is obtained for
$\alpha=0.01$, $0.05$, and $0.10$. The shared-LS ablation shows that
the uncertainty-aware clusters allow multiple working conditions to
share a common model. Sparse estimation further reduces the number of
active parameters by 24.7\%--50.0\% without materially changing the
average validation error. In comparison, UACS reduces the average MSE by approximately 43.4\% relative to K-means-based grouping, indicating the benefit of accounting for parameter uncertainty when constructing shared models across working conditions.

\section{Conclusion}

This paper developed a unified uncertainty-aware and sparsity-promoting framework for WCF estimation under diverse and unlabeled working conditions. By integrating uncertainty-aware similarity assessment, unsupervised clustering, and sparse parameter estimation into a single framework, the proposed method enables reliable model comparison, effective condition clustering, and more parsimonious parameter representations.

Experimental results on real vehicle data demonstrated that the proposed framework achieves prediction accuracy comparable to condition-specific LS models while significantly reducing both the number of models and the number of active parameters. In addition, it consistently outperforms conventional distance-based clustering methods such as K-means, highlighting the importance of accounting for parameter uncertainty in model grouping.

Overall, the results showed that jointly incorporating parameter uncertainty, condition grouping, and sparse parameter estimation enables effective model reduction in multi-condition system identification. The proposed framework offers a practical approach for constructing efficient and interpretable models under varying working conditions. Future work will focus on extending the proposed framework to nonlinear modeling settings and investigating adaptive implementations for real-time applications.

\section*{Acknowledgment}

We sincerely appreciate Mr. Yu Qiu from the SAIC Motor R\&D Innovation Headquarters for providing the real-world data collected from the prototype vehicle.

\bibliography{ref}
\bibliographystyle{IEEEtran}

\end{document}